\documentclass[twocolumn,showpacs,preprintnumbers,amsmath,amssymb,prl]{revtex4}

\usepackage{graphicx}
\usepackage{dcolumn}
\usepackage{bm}% bold math

\usepackage{mathptmx, courier, pifont}
\usepackage[scaled=0.92]{helvet}
\usepackage[T1]{fontenc}
\usepackage{textcomp}
\usepackage{color}
\usepackage{colordvi}

\newcommand{\quarterthin}{\kern 0.0417em}

\begin{document}

%\preprint{APS/123-QED}

\title{Isoscalar neutron-proton pairing and SU(4)-symmetry breaking
in Gamow-Teller transitions}

\author{K.~Kaneko$^{1}$, Y.~Sun$^{2,3,4}$, T.~Mizusaki$^{5}$}

\affiliation{
$^{1}$Department of Physics, Kyushu Sangyo University, Fukuoka
813-8503, Japan \\
$^{2}$School of Physics and Astronomy, Shanghai Jiao Tong
University, Shanghai 200240, China \\
$^{3}$Collaborative Innovation Center of IFSA, Shanghai
Jiao Tong University, Shanghai 200240, China \\
$^{4}$Institute of Modern Physics, Chinese Academy of Sciences,
Lanzhou 730000, China \\
$^{5}$Institute of Natural Sciences, Senshu University, Tokyo
101-8425, Japan
}

\date{\today}

\begin{abstract}
The isoscalar neutron-proton pairing is thought to be important for
nuclei with equal number of protons and neutrons but its
manifestation in structure properties remains to be understood. We
investigate the Gamow-Teller (GT) transitions for the
$f_{7/2}$-shell nuclei in large-scale shell-model calculations with
the realistic Hamiltonian. We show that the isoscalar $T=0,
J^{\pi}=1^{+}$ neutron-proton pairing interaction plays a decisive
role for the concentration of GT strengths at the first-excited
$1_{1}^{+}$ state in $^{42}$Sc, and that the suppression of these
strengths in $^{46}$V, $^{50}$Mn, and $^{54}$Co is mainly caused by
the spin-orbit force  supplemented by the quadrupole-quadrupole
interaction. Based on the good reproduction of the charge-exchange
reaction data, we further analyze the interplay between the
isoscalar and isovector pairing correlations. We conclude that even
for the most promising $A=42$ nuclei where the SU(4)
isoscalar-isovector-pairing symmetry is less broken, the probability
of forming an isoscalar neutron-proton pairing condensation is less
than 60\% as compared to the expectation at the SU(4)-symmetry
limit.
\end{abstract}


\maketitle

%=================================================================
{\it Introduction.} Pairing of two kinds of fermions is a unique
phenomenon in nuclear physics. According to Heisenberg
\cite{Heisenberg32}, neutrons and protons can be
regarded as two "states" of nucleons, described by isospin $T$.
Thus, nuclear many-body states and interactions are characterized
by combinations of spin and isospin \cite{Osterfeld92}, respecting the
antisymmetrization requirement for wave functions. For most nuclei
with $N> Z$, the isovector ($T=1$) neutron-neutron ($nn$) and
proton-proton ($pp$) pairings coupled to angular momentum zero
($J=0$) can be well described by the BCS-type models
\cite{BCS57,BMP58,Be59} similarly as in condensed-matter physics.
For nuclei near the $N=Z$ line, the neutrons and protons close to
the Fermi surface occupy identical orbits, and thus can have large
probabilities to form isovector neutron-proton ($np$) pairs with
$T=1$ or isoscalar ($T=0$) $np$ pairs. With a short-range force, the
angular momentum of these $np$ pairs favors $J=0$ for $T=1$, and
$J=1$ or $J=J_{max}$ for $T=0$ \cite{FM14}. It has been shown that
the $np$ pairing with $T=1, J=0$ should be treated on an equal
footing with $T=1, J=0$ $nn$ and $pp$ pairing \cite{FS99}.

The isoscalar $T=0$ $np$ pairing, on the other hand, is special in
nuclear physics, and has long been a discussion focus. An
interesting question has been whether the $T=0$ $np$-pairing
condensate can occur in nuclei. $\beta$-decay and charge-exchange
reaction, which involve isospin-flip, are thought
to be the means to study the $T=0$ $np$-pairing~\cite{FM14}. In the
recent ($^3$He, $t$) charge-exchange experiments, Fujita {\it et
al.} \cite{Fujita14,Fujita15} have shown an interesting observation
in the $^{42}$Ca $\rightarrow$ $^{42}$Sc reaction. They found that,
instead of the usual expectation for nuclei with mass $A>60$ that
most of Gamow-Teller (GT) strengths are distributed with a few MeV
width at $E_x>9$ MeV \cite{RS94}, the GT strengths they obtained
concentrate in the lowest excited $1^{+}$ state at 0.6 MeV in
$^{42}$Sc, which they call the low-energy super GT state. Moreover,
in the same reaction for all other $f_{7/2}$-shell nuclei, which
ends up with the odd-odd $N=Z$ ($T_z=0$) nuclei (i.e., $^{46}$V,
$^{50}$Mn, and $^{54}$Co), the GT strength distributions are found
to be qualitatively different. As mass number $A$ increases, the
low-energy strength becomes fragmented and the bumplike structure in
the high-energy region begins to develop. Finally in $^{54}$Co, the
distribution is mainly of the typical bumplike GT resonances. These
results seem to suggest drastic differences occurring along the
$N=Z$ line, which have attracted our attention.

Combining the $T=0,J^{\pi}=1^{+}$ mode with the $T=1,J^{\pi}=0^{+}$
one, the isovector and isoscalar pairing correlations have been
examined by the exactly solvable SO(8) model with degenerate
single-particle orbits \cite{Evans81,Engel97}. In the SU(4) symmetry
limit, the GT strength is, as pointed out by Wigner \cite{Wigner37},
indeed expected to concentrate at the lowest $1^{+}$ state. Thus the
large GT strength at the lowest $1^{+}$ state observed in $^{42}$Sc
\cite{Fujita14,Fujita15} would correspond to an SU(4) symmetry, and
it could be the fingerprint of the $T=0$ pairing. The GT strength
distributions with $A=42-58$ were studied by using a self-consistent
Skyrme Hartree-Fock-Bogoliubov plus quasiparticle random-phase
approximation (QRPA) method including isoscalar and isovector
residual interactions \cite{Bai14}. In Ref. \cite{Yoshida14},
Yoshida considered the lowest $1_{1}^{+}$ in $^{42}$Sc as a
precursory soft mode of the $T=0$ pairing condensation. There was
reported experiment for the GT decay to the odd-odd $N=Z$ $^{62}$Ga
suggesting no evidence of the isoscalar pairing condensation
\cite{Grodner11}. It is important to mention that the SU(4) symmetry
can be largely suppressed due to the presence of the spin-orbit (SO)
splitting \cite{Poves98,Bertsch10}. The isoscalar spin-triplet
pairing interaction is significant for the enhancement of the GT
strength, while the experimental evidence of the $T=0$ pairing
condensation is still controversial \cite{FM14,Machiavelli00}.

Thus, despite the great efforts made in the past, both
experimentally and theoretically, there has been no consensus on the
$T=0$ $np$-pairing condensation in nuclei. A realistic model that
describes the shell effect and contains all kinds of relevant
interactions, which either favor or unfavor the $np$-pairing
condensate, could shed light on this question.

{\it The PMMU model.} The present authors have recently proposed a
unified realistic shell-model Hamiltonian called PMMU
\cite{Kaneko14,Kaneko15}, consisting of the two-body interactions in
separable forms with the monopole interaction $V_{m}^{MU}$
constructed from the monopole-based universal force
\cite{Otsuka10b},
\begin{eqnarray}
 H & = & H_{0} + H_{PM}  + H_{m}^{MU},  \\
 H_{0} & = & \sum_{\alpha} \varepsilon_a c_\alpha^\dag c_\alpha, \\
 H_{PM}  & = &
 -  \frac{1}{2} \sum_{J=0,2} g_J \sum_{M\kappa} P^\dag_{JM1\kappa} P_{JM1\kappa}
 -  \frac{1}{2} \chi_2 \sum_M :Q^\dag_{2M} Q_{2M}: \nonumber \\
   && - \frac{1}{2} g_1 \sum_{M} P^\dag_{1M00} P_{1M00} \\
 H_{m}^{MU}  & = &  \sum_{a \leq b, T} V_{m}^{MU}(ab,T)
 \sum_{JMK}A^\dagger_{JMTK}(ab) A_{JMTK}(ab).
\end{eqnarray}
It was demonstrated \cite{Kaneko14} that the PMMU interaction can
well describe nuclear properties of the $pf$ and $pf_{5/2}g_{9/2}$
shell nuclei. In Eq. (3), the first term denotes the isovector $J=0$
pairing ($P01$) and $J=2$ pairing ($P21$) interactions in the
particle-particle channel, and the second term the
quadrupole-quadrupole ($QQ$) interaction in the particle-hole
channel. The isoscalar $J=1$ pairing ($P10$) interaction in the
particle-particle $T=0$ channel is added into the PMMU Hamiltonian
as the third term in Eq. (3). The monopole interaction $H_{m}^{MU}$
is constructed from the monopole-based universal force
\cite{Otsuka10b}, and its effect has been extensively discussed in
Refs. \cite{Kaneko14,Kaneko15}.

In the present work, we use the same single-particle energies
$\varepsilon_a$ and the parameter strengths of our previous paper
\cite{Kaneko14} for the $pf$ shell space. The spin-orbit splitting
6.5 MeV between the spin-orbit parter $f_{7/2}$ and $f_{5/2}$ is
chosen so as to fit the measured value in $^{41}$Ca. The isoscalar
monopole terms are slightly shifted so as to fit the observed lowest
$1_{1}^{+}$ energies for $^{46}$V, $^{50}$Mn, and $^{54}$Co. For the
added $T=0,J=1$ pairing interaction, the force strength is given by
$g_{1}=g_{1}^{0}/A$, which has the same mass-dependence as the
$T=1,J=0$ interaction. It should be mentioned that there is no
common agreement for the choice of $g_{1}$
\cite{Yoshida14,Hinohara14}. We have confirmed that this additional
$P10$ force does not change the results and conclusions obtained in
our previous paper \cite{Kaneko14}. Shell-model calculations are
performed with the code MSHELL64 \cite{Mizusaki}.

%===============  fig. 1  ========================================
\begin{figure}[t]
\includegraphics[totalheight=4.8cm]{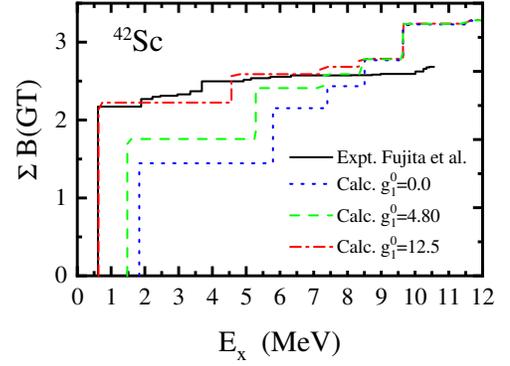}
\caption{(Color online) Cumulative sums of B(GT) in $^{42}$Sc,
calculated by the PMMU Hamiltonian including the $T=0,J=1$ interaction
with strengths $g_{1}^{0}$ = 0.0, 4.80, and 12.5, and compared with
the experimental data from Ref. \cite{Fujita15}.}
 \label{fig1}
\end{figure}
%=================================================================

%===============  fig. 2 ========================================
\begin{figure}[t]
\begin{center}
\includegraphics[totalheight=4.5cm]{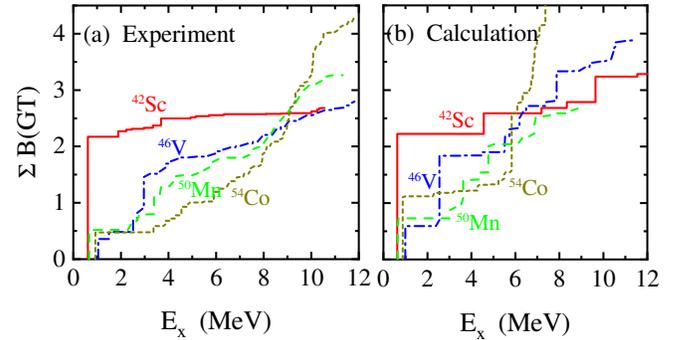}
\caption{(Color online) Cumulative sums of B$(GT)$ of
the daughter nuclei $^{42}$Sc, $^{46}$V, $^{50}$Mn, and $^{54}$Co
for (a) experimental data \cite{Fujita15} and (b) calculation.}
  \label{fig2}
 \end{center}
\end{figure}
%=================================================================

{\it On the SU(4)-symmetry breaking.} Figure \ref{fig1} shows the
calculated B$(GT)$ distributions and the experimental results in the
$^{42}$Ca $\rightarrow$ $^{42}$Sc reaction. The quenching factor $q$
= 0.74 is used for the GT-strength calculation as it proves to be
appropriate for the mass region of $A\sim 40-60$ \cite{Martinez96}.
We can see that the $P10$ pairing interaction with the parameter
$g_{1}^{0}$ = 12.5 MeV reproduces the B$(GT)$ data as well as the
lowest $1_{1}^{+}$ excitation energy. With weaker $g_{1}^{0}$, the
GT strength is suppressed, and the lowest $1^{+}$ excitation is
shifted to higher energies. This demonstrates clearly that a
sufficiently strong isoscalar $P10$ pairing interaction is essential
for the occurrence of the low-energy super GT state observed in
Refs. \cite{Fujita14,Fujita15}. The obtained sudden increase of the
GT strength at $E_x=9.64$ MeV for all calculations with different
$P10$ pairing strengths is due to the $T=1$ states. Previously the
shell-model calculation using the GXPF1J interaction \cite{Honma05}
was performed to study the same data \cite{Fujita14,Fujita15}. The
predicted $1_{1}^{+}$ excitation energy 0.33 MeV was about half of
the experimental energy 0.611 MeV, while the GT distribution was
reproduced well.

Figure \ref{fig2}(a) summarizes the experimental B$(GT)$
distributions \cite{Fujita15} as functions of excitation energies of
the daughter nuclei $^{42}$Sc, $^{46}$V, $^{50}$Mn, and $^{54}$Co,
in which strong mass-dependence is evident. The GT strength
concentrates in the lowest $1_{1}^{+}$ state of $^{42}$Sc, but moves
to higher energies with more spread patterns for $^{46}$V,
$^{50}$Mn, and $^{54}$Co. The calculated results in Fig.
\ref{fig2}(b) show qualitatively the same trend.

Our employed shell-model Hamiltonian in Eq. (1) is rich in physics,
which contains all relevant interactions in the form of separable
forces. The $H_0$ term and the monopole interaction $H_{m}^{MU}$
produce the realistic single-particle states, ensuring that the SO
effect is  correctly described. In the particle-particle channel,
the $P10$, $P21$, and $P01$ terms present the competition of
isoscalar and isovector pairing, and in the particle-hole channel,
$QQ$ describes the correlation induced by deformation. If we assume
the degenerate single-particle orbits and take only the $T=1,J=0$
and $T=0,J=1$ pairing interactions in the PMMU, it reduces to the
solvable SO(8) model \cite{Evans81,Engel97}. Further, if the $T=0$
and $T=1$ pairing force strengths are set to be equal, the
Hamiltonian is invariant under SU(4), in which the GT strength to
the lowest $1^{+}$ state becomes large \cite{Wigner37}. In turn, the
appearance of the large GT strength at the lowest $1^{+}$ state
implies that both pairing correlations are equally significant.
However, it has been discussed that this large GT value is
suppressed due to strong SO splitting \cite{Bertsch10}. Thus it is
very interesting to study the contribution of various terms in the
realistic PMMU Hamiltonian to the GT strengths, to answer the
question whether and under which conditions the isoscalar $np$
pairing can lead to a collective mode coexisting with that of the
isovector pairing.

\begin{figure}[t]
 \begin{center}
\includegraphics[totalheight=4.8cm]{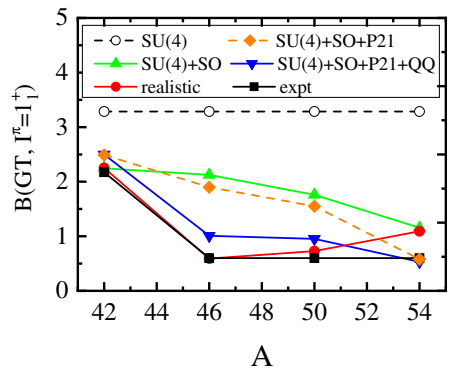}
\caption{(Color online) Calculated B$(GT)$ strengths of
the lowest $1_{1}^{+}$ states in the daughter nuclei $^{42}$Sc,
$^{46}$V, $^{50}$Mn, and $^{54}$Co with separate contributions from
each interaction in Eqs. (2)-(4), compared with the experimental
data \cite{Fujita14} and the value of the SU(4) symmetry. }
  \label{fig3}
 \end{center}
\end{figure}

In Fig. \ref{fig3}, the effect of each term in Eq. (1) is studied
with the B$(GT)$ for the first excited $1_{1}^{+}$ states in the
$A=42-54$ daughter nuclei. If $QQ$ and $P21$ in Eq. (3) and the
monopole $H_{m}^{MU}$ interaction in Eq. (4) are all switched off
and the SO splitting is excluded from the calculation, Eq. (1)
becomes the Hamiltonian with only the isoscalar $P10$ and isovector
$P01$ pairing terms. Further, if the strengths of both pairing terms
are set to be equal, i.e., $g_{1}=g_{0}$, then it is invariant under
SU(4). In the limit of the exact SU(4) symmetry, the GT strength is
concentrated in a large single transition to the lowest $1_{1}^{+}$
GT state \cite{Wigner37}. Since the second term in the Ikeda sum
rule for the GT strength, $\sum B(GT_{-})-\sum
B(GT_{+})=3(N-Z)q^{2}$, can be neglected, the SU(4) symmetry
corresponds to a constant GT strength $B(GT)=6q^{2}\approx 3.29$ for
all cases of $A=42-54$, as shown in Fig. \ref{fig3}.

The departure from the SU(4) symmetry limit occurs as these terms
are put back. The SO splitting in the $pf$-shell is known to be
sizable, giving rise to the magic number 28. The realistic SO force
leads to the SU(4) symmetry breaking and suppression of the GT
strengths. As seen in Fig. \ref{fig3}, the largest SO effect is
found at $A=54$, where about 2/3 of the SU(4) value is suppressed.
Among the four cases, $A=42$ has the smallest influence. On the
other hand, the inclusion of the $J=2$ pairing does not bring much
effect to GT. However, when the $QQ$ interaction is switched on, the
B$(GT)$ for $A=46, 50$ are drastically reduced, while the effect on
$A=42, 54$ is smaller. Thus the $QQ$ interaction is largely
responsible for the small B$(GT)$ values for $A=$46 and 50. With the
monopole interactions included and $g_{1}^{0}$ = 0.26, finally the
realistic PMMU Hamiltonian reproduces very well the experimental
data. Only for $A=54$, the calculated result is somewhat larger than
the data.

%===============  fig. 3 ========================================
%=================================================================

{\it Question on the isoscalar neutron-proton pairing condensation.}
One important finding in Fig. \ref{fig3} is that the B($GT$) of
$A=42$ has the weakest breaking of the SU(4) symmetry, and it is
caused mainly by the SO splitting. In an even-even nucleus with the
isovector $J = 0$ pairing interaction, the ground state with
seniority zero is usually interpreted as a condensate of isovector
$J=0$ pairs. Then the isoscalar pairing condensate may be realized
for the isoscalar $J = 1$ pairing Hamiltonian with the degenerate
single-particle orbits. However, this isoscalar pairing condensate
may not persist against the single-particle splitting due to the SO
force \cite{Bertsch10}. With all these factors taking into account,
we formulate an isoscalar-isovector-pairing Hamiltonian
\begin{eqnarray}
 H & = & H_{0}
  -  \frac{g(1-x)}{2}  \sum_{\kappa} P^\dag_{001\kappa} P_{001\kappa}
  -  \frac{g(1+x)}{2}  \sum_{M} P^\dag_{1M00} P_{1M00}, \nonumber \\ \label{simpleH}
\end{eqnarray}
with $x$ being a control parameter. Equation (\ref{simpleH}) can be
regarded as a simplified model from the realistic PMMU Hamiltonian
(1) to study the competition of isoscalar and isovector pairings
under the presence of the SO splitting. If the single-particle
energies in $H_{0}$ are degenerate, the Hamiltonian (5) is invariant
under SO(8), and $x=-1$ and $x=1$ corresponds, respectively, to the
usual isovector $J=0$ pairing and isoscalar $J=1$ pairing
Hamiltonian. For $x=0$, the Hamiltonian (\ref{simpleH}) is invariant
under SU(4). Thus by varying $x$, it describes the phase evolution
from the $T=1, J=0$ pairing phase ($x=-1$), through the SU(4) Wigner
super-multiplet phase ($x=0$), to the $T=0, J=1$ pairing phase
($x=1$).

%===============  fig. 4 ========================================
\begin{figure}[t]
\includegraphics[totalheight=7.6cm]{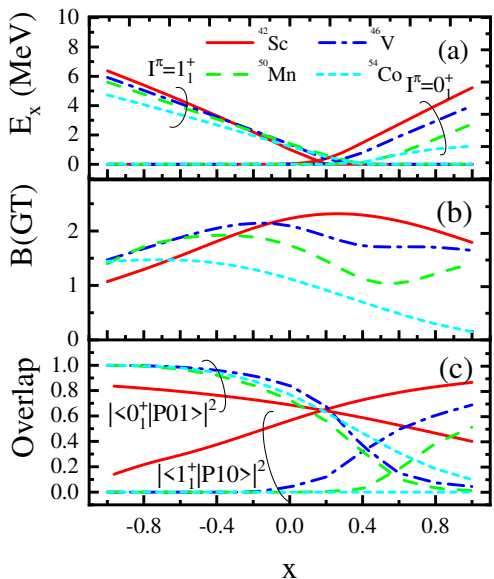}
\caption{(Color online) Excitation energy, B$(GT)$ value, and
overlap between the wavefunctions by solving Eq. (\ref{simpleH}) and
the realistic Hamiltonian (1), as functions of the control parameter
$x$.}
  \label{fig4}
\end{figure}
%=================================================================

Figure \ref{fig4} shows the calculated results obtained by solving
the Hamiltonian (\ref{simpleH}), where the non-degenerate realistic
single-particle energies \cite{Kaneko14} are employed. The
competition between the lowest $0_{1}^{+}$ and $1_{1}^{+}$ states to
be the ground state are shown in Fig. \ref{fig4}(a) for the $T_z=0$
nuclei $^{42}$Sc, $^{46}$V, $^{50}$Mn, and $^{54}$Co. It is seen
that for all these odd-odd $N=Z$ nuclei, $0_{1}^{+}$ is definitely
the lowest state for all negative $x$'s and small positive $x$'s.
Starting from $x=-1$, all the $1_{1}^{+}$ energies drop continuously
with increasing $x$, and are inverted in position with the
$0_{1}^{+}$ states at a critical point around $x_c=0.20$, after
which the $0_{1}^{+}$ energies increase. For $^{42}$Sc, $x_c=0.17$.
A correlated result is shown in Fig. \ref{fig4}(b), where from
$x=-1$, the calculated B$(GT)$ transition from $^{42}$Ca to the
daughter $^{42}$Sc increases with $x$, and reaches the maximum at
$x_{c}=0.17$.

There has been a long-standing question whether a strong isoscalar
pairing correlation can lead to condensation. The possibility of
finding the isoscalar pairing condensate was theoretically
investigated for heavy nuclei \cite{Bertsch10}. We note that the
Hamiltonian in the present work is superior to that used in Ref.
\cite{Bertsch10}. To see what our shell model calculation may
suggest, we follow the definition for the isovector and isoscalar
pairing condensates in Refs. \cite{Dufour96,Poves98}. The isovector
$J=0$ pairing condensation state $|P01\rangle$ is obtained by
solving the second term and the first term with experimental
single-particle energies in Eq. (\ref{simpleH}). The isoscalar $J=1$
pairing condensate state $|P10\rangle$ is obtained by solving the
third term and the first term with degenerate single-particle
energies. The probabilities of finding a nuclear state in different
condensed phases are given by the squared overlap of that state with
$|P01\rangle$ or $|P10\rangle$.

In Fig. \ref{fig4}(c), the squared overlaps $|\langle
0_{1}^{+}|P01\rangle |^{2}$ and $|\langle 1_{1}^{+}|P10\rangle
|^{2}$ are shown as functions of $x$ in Eq. (\ref{simpleH}). For all
the four $f_{7/2}$-shell odd-odd $N=Z$ nuclei, $|\langle
0_{1}^{+}|P01\rangle |^{2}$ is overwhelmingly large when $x$ is
negative. Beyond $x=0$, it begins to decrease with positive $x$'s.
At $x=+1$, this quantity is close to zero except for $^{42}$Sc. On
the other hand, the probability of finding the isoscalar pair
condensate in the $1_{1}^{+}$ state of $^{46}$V, $^{50}$Mn, and
$^{54}$Co is zero for the negative $x$'s. As $x$ varies from $-1$ to
$+1$, the overlap $|\langle 1_{1}^{+}|P10\rangle |^{2}$ for
$^{42}$Sc increases monotonously with no drastic changes, and
$|1_{1}^{+}\rangle$ evolves smoothly from the isovector into the
isoscalar pairing phase. At $x=0$ (the SU(4) symmetry limit), the
overlaps of the condensation states with the physical ground
$0_{1}^{+}$ and excited $1_{1}^{+}$ states of $^{42}$Sc are obtained
as $|\langle 0_{1}^{+}|P01\rangle |^{2}$ = 0.99 and $|\langle
1_{1}^{+}|P10\rangle |^{2}$ = 0.57, respectively.

In sharp contrast, the overlap $|\langle 1_{1}^{+}|P10\rangle |^{2}$
for $^{46}$V and $^{50}$Mn begins to take a nonzero value only when
$x$ is positive, and for $^{54}$Co, $|\langle 1_{1}^{+}|P10\rangle
|^{2}$ is always zero for any positive $x$. In particular, with the
realistic parameter around $x=0$ corresponding to the SU(4)
symmetry, $|\langle 1_{1}^{+}|P10\rangle |^{2}$ is zero for
$^{46}$V, $^{50}$Mn, and $^{54}$Co. This means that for the four
cases in the $f_{7/2}$-shell, it is impossible to realize an
isoscalar pairing condensation. In $^{42}$Sc, the calculated
probability $|\langle 1_{1}^{+}|P10\rangle |^{2}$ is 57\%. The
formed isoscalar $J=1$ pairing coexists with the isovector $J=0$
pairing, while $^{42}$Sc contains only one valence $np$ pair.

{\it Summary.} In conclusion, to investigate the role of the
isoscalar $np$ pairing interaction in GT transitions and the
possibility of $np$ pairing condensation in nuclei, we performed
large-scale shell-model calculations with the realistic PMMU
Hamiltonian for the $f_{7/2}$-shell nuclei. The
early works based on simpler models \cite{Cha83,Spain03} are close
in spirit to the present work. The isoscalar $T=0, J=1$ interaction
is found to be decisively important for explaining the large B$(GT)$
strengths of the transition from the ground $0_{1}^{+}$ state in
$^{42}$Ca to the lowest $1_{1}^{+}$ state in $^{42}$Sc. The
systematics of B$(GT)$ distributions are well described for the mass
$A=42, 48, 50, 54$. Our realistic shell-model calculation
with more realistic interactions confirms the
previous conclusion \cite{Bertsch10} that the $B(GT)$ strength in
these nuclei is considerably suppressed by the SO splitting and the
SU(4) symmetry is broken. We also show that the $B(GT)$ strengths
are largely suppressed by the $QQ$ interaction for the masses $A=46$
and 50. The deformation effect in GT transitions
was emphasized in Ref. \cite{Johnson04}. We discussed the isovector
and isoscalar pairing condensation using a simplified model reduced
from the PMMU Hamiltonian. The results suggest that even for the
most promising $A = 42$ nuclei where the SU(4)
isoscalar-isovector-pairing symmetry is less broken, the probability
of forming an isoscalar $np$ pairing condensate is less than 60\% as
compared to the idealized situation. Finally, we
note that the present realistic shell-model calculation may be
applied to the study of SU(4)-symmetry in double-$\beta$ decay, as
discussed recently in Ref. \cite{Ferreira17}.

{\it Acknowledgement.} One of us (YS) thanks Y.
Fujita for valuable discussions during his visit in Shanghai. This
work was partially supported by the National Natural Science
Foundation of China (No. 11575112) and by the National Key Program
for S\&T Research and Development (No. 2016YFA0400501).

%------------------------------------------------------------------------------

%------------------------------------------------------------------------------

\end{document}